\documentclass[11pt]{article}
\usepackage[utf8]{inputenc}
\usepackage{amsmath, amssymb, amsfonts}
\usepackage{authblk}
\usepackage{graphicx} 
\usepackage{braket}    
\usepackage[margin=1in]{geometry}
\usepackage{longtable}
\usepackage{array}
\usepackage{makecell}
\usepackage{float}
\usepackage{multirow}
\usepackage[section]{placeins}
\usepackage{caption}
\usepackage{booktabs}
\usepackage{subcaption}

\title{\textbf{Mentor-Initiated Asymmetric Bidirectional Quantum Teleportation Protocol for Arbitrary Qubit States}}

\author[1]{Sugata Adhya\thanks{Corresponding author: sugataadhya@yahoo.com}}
\author[2]{Benchao Yang \thanks{yang\_benchao@aliyun.com}}

\affil[1]{The Bhawanipur Education Society College, Kolkata, India}
\affil[2]{Information Engineering University, Zhengzhou, China}

\date{}

\begin{document}

\maketitle

\begin{abstract}
In this paper, a mentor-initiated bidirectional asymmetric quantum teleportation protocol is proposed. Unlike existing schemes that restrict the two-qubit state transmitted from Bob to Alice to a Bell-like form $\alpha|00\rangle+\beta|11\rangle$, our protocol allows an arbitrary two-qubit state to be teleported from Bob to Alice within the mentor-initiated framework. The required entangled resources are constructed using standard quantum gates, and the protocol is illustrated through its quantum-circuit implementation in Qiskit. The proposed protocol is further simulated using Qiskit, and the simulation results show probability distributions very close to the theoretically expected values for the chosen input states.
\end{abstract}

\vspace{0.5em}

\section{Introduction}

Quantum teleportation enables the transfer of quantum information by exploiting the property of quantum entanglement, providing a powerful framework for high-capacity and secure quantum communication compared with classical communication. The process utilizes shared entangled resources and classical communication to transfer an unknown quantum state from a sender to a receiver without physically transmitting the state itself.

In 1993, Bennett et al. \cite{Bennett1993} first proposed a quantum teleportation protocol employing an Einstein--Podolsky--Rosen (EPR) entangled pair as the quantum resource for transferring an unknown quantum state between two parties. This pioneering result opened up possibilities for the development of numerous quantum protocols for different forms of quantum teleportation, significantly advancing the field of quantum communication. For further details, interested readers may refer to \cite{Choudhury2026} and the references therein.

Among the various schemes for enhanced quantum teleportation, bidirectional quantum teleportation protocols provide a framework for the simultaneous exchange of quantum information between two parties. Bidirectional teleportation was subsequently investigated in different forms, including asymmetric protocols in which the numbers of qubits transmitted in the two directions are different. Following these developments, several bidirectional quantum teleportation protocols have been proposed using different entangled quantum resources and communication schemes. For example, interested readers may consult \cite{Aisan2025,Choudhury2023b,Hong2016,mentorBQT,Wang2022,Yang2016,Zha2013,Zhang2015} and the references therein.

Within the class of bidirectional teleportation schemes, mentor-initiated protocols provide an additional mechanism for controlling the establishment of the quantum channel. In a mentor-initiated scheme, the mentor is a party who initiates the protocol by participating in the creation of the entangled channel between the sender and the receiver. Initially, the sender and receiver do not directly share the required entangled channel; rather, the mentor participates in establishing the entanglement needed for subsequent quantum communication. Once the entangled channel has been established, the mentor's role in the protocol is completed, allowing the sender and receiver to perform quantum teleportation through the established channel. Mentor-initiated schemes have subsequently been investigated in various forms, including remote state preparation, bidirectional controlled teleportation, and bidirectional asymmetric teleportation \cite{Choudhury2019,Choudhury2022,Choudhury2023a,Choudhury2023b,Jing2024a,mentorBQT}.

In particular, the authors of \cite{mentorBQT} proposed a mentor-initiated controlled bidirectional asymmetric quantum teleportation protocol in which Alice teleports an arbitrary one-qubit state to Bob, whereas Bob teleports a two-qubit Bell-like state of the form $\alpha|00\rangle+\beta|11\rangle$ to Alice. The protocol is initiated by a mentor and operates in the presence of controllers. However, their scheme is not applicable to the teleportation of an arbitrary two-qubit state from Bob to Alice. In our work, we address this limitation by proposing a mentor-initiated bidirectional asymmetric quantum teleportation protocol that enables the teleportation of an arbitrary one-qubit state from Alice to Bob and an arbitrary two-qubit state from Bob to Alice. We construct the required entangled resources and demonstrate the proposed protocol through its quantum-circuit implementation and simulation using Qiskit.

This paper is organized as follows. In Section \ref{sec2}, we propose an arbitrary \(1\leftrightarrow2\)-qubit asymmetric bidirectional quantum communication protocol, initiated by a mentor. Section \ref{sec3} is devoted to the construction of the entangled resources required for the proposed protocol using Hadamard, CNOT, and SWAP gates. In Section \ref{sec4}, we present a practical simulation of the proposed protocol using Qiskit and illustrate its operation through representative examples. Section \ref{sec5} is devoted to the efficiency analysis of the proposed protocol. Finally, Section \ref{sec6} concludes the paper with a discussion of the outcomes obtained from the study.

\section{Proposed Protocol}\label{sec2}

In this protocol, Alice wants to transmit an arbitrary one-qubit state to Bob. At the same time, Bob wants to transmit an arbitrary two-qubit state to Alice. Initially, Alice and Bob do not share any entangled quantum channel between them. However, there is a mentor who is connected to both Alice and Bob through two different entangled channels.

Let the one-qubit state that Alice wants to transmit to Bob be

$$|\psi_1\rangle_{a_1}=x|0\rangle+y|1\rangle$$

and the two-qubit state that Bob wants to send to Alice be 

$$|\psi_2\rangle_{b_1b_2}=\alpha|00\rangle+\beta|01\rangle+\gamma|10\rangle+\delta|11\rangle$$

where $x,y,\alpha,\beta,\gamma,\delta\in\mathbb C$ such that $|x|^2+|y|^2=1$ and $|\alpha|^2+|\beta|^2+|\gamma|^2+|\delta|^2=1.$

We consider two entangled quantum resources $\left|\Omega_{1}\right\rangle_{M-A}$ shared between mentor and Alice, and $\left|\Omega_{2}\right\rangle_{M-B}$ shared between mentor and Bob, where

$$\begin{aligned}
|\Omega_1\rangle_{M-A} &= \frac{1}{2\sqrt{2}} \Big( |000000\rangle + 
|001100\rangle + |010010\rangle + |011110\rangle \\
& \qquad \qquad \qquad \qquad \qquad + |100001\rangle + |101101\rangle + |110011\rangle + |111111\rangle \Big)_{m_1 m_2 m_3 A_1 A_2 A_3} \\
|\Omega_2\rangle_{M-B} &= \frac{1}{2\sqrt{2}} \Big( |000000\rangle + |001001\rangle + |010010\rangle + |011011\rangle \\
& \qquad \qquad \qquad \qquad \qquad + |100100\rangle + |101101\rangle + |110110\rangle + |111111\rangle\Big)_{m_4 m_5 m_6 B_1 B_2 B_3} \\
\end{aligned}$$

where the qubits $m_1,m_2,m_3,$ and $m_4,m_5,m_6$ are held by the mentor $M$; $A_1,A_2,A_3$ by Alice ($A$); $B_1,B_2,B_3$ by Bob ($B$).

The overall entangled channel is represented by
$
|\phi\rangle
=
|\Omega_{1}\rangle_{M-A}
\otimes
|\Omega_{2}\rangle_{M-B},
$
while the initial state of the complete system is
$
|\chi\rangle
=
|\phi\rangle
\otimes
|\psi_{1}\rangle_{a_{1}}
\otimes
|\psi_{2}\rangle_{b_{1}b_{2}}.$

The mentor initiates the protocol by performing Bell-basis measurements simultaneously on the qubit pairs $(m_1,m_4),$ $(m_2,m_5),$ and $(m_3,m_6)$. Here, the four maximally entangled two-qubit Bell states are defined as
\begin{equation}
|\phi^{\pm}\rangle = \frac{1}{\sqrt{2}}\Big(|00\rangle \pm |11\rangle\Big), \qquad 
|\psi^{\pm}\rangle = \frac{1}{\sqrt{2}}\Big(|01\rangle \pm |10\rangle\Big).
\end{equation} 

The mentor communicates the outcomes of the Bell-basis measurements $(m_1,m_4)$ and $(m_2,m_5)$ to Alice and $(m_3,m_6)$ to Bob through classical communication channels. At this stage, the mentor's role in the protocol is complete. Following the Bell-basis measurements performed on the qubit pairs $(m_1,m_4),$ $(m_2,m_5),$ and $(m_3,m_6)$, the composite state of the remaining qubits $A_1A_2A_3B_1B_2B_3$ collapses into one of the $64$ possible conditional states, depending on the corresponding measurement outcomes. The $64$ conditional states
$|\xi_{\alpha\beta\gamma}\rangle_{A_1A_2A_3B_1B_2B_3}$
corresponding to the Bell-state measurement outcomes
$\alpha$, $\beta$, and $\gamma$ are given by the following matrix equation:

$$
\begin{pmatrix}
|\xi_{\alpha^+\beta^+\gamma^+}\rangle\\
|\xi_{\alpha^+\beta^+\gamma^-}\rangle\\
|\xi_{\alpha^+\beta^-\gamma^+}\rangle\\
|\xi_{\alpha^+\beta^-\gamma^-}\rangle\\
|\xi_{\alpha^-\beta^+\gamma^+}\rangle\\
|\xi_{\alpha^-\beta^+\gamma^-}\rangle\\
|\xi_{\alpha^-\beta^-\gamma^+}\rangle\\
|\xi_{\alpha^-\beta^-\gamma^-}\rangle
\end{pmatrix}
=
\frac{1}{2\sqrt{2}}
\begin{pmatrix}
+1 & +1 & +1 & +1 & +1 & +1 & +1 & +1\\
+1 & -1 & +1 & -1 & +1 & -1 & +1 & -1\\
+1 & +1 & -1 & -1 & +1 & +1 & -1 & -1\\
+1 & -1 & -1 & +1 & +1 & -1 & -1 & +1\\
+1 & +1 & +1 & +1 & -1 & -1 & -1 & -1\\
+1 & -1 & +1 & -1 & -1 & +1 & -1 & +1\\
+1 & +1 & -1 & -1 & -1 & -1 & +1 & +1\\
+1 & -1 & -1 & +1 & -1 & +1 & +1 & -1
\end{pmatrix}
\begin{pmatrix}
|A^1_{\alpha\beta\gamma}\rangle\\
|A^2_{\alpha\beta\gamma}\rangle\\
|A^3_{\alpha\beta\gamma}\rangle\\
|A^4_{\alpha\beta\gamma}\rangle\\
|A^5_{\alpha\beta\gamma}\rangle\\
|A^6_{\alpha\beta\gamma}\rangle\\
|A^7_{\alpha\beta\gamma}\rangle\\
|A^8_{\alpha\beta\gamma}\rangle
\end{pmatrix}.
$$ The states $|A^i_{\alpha\beta\gamma}\rangle$ are given in Table~\ref{tab:conditional}.

\begin{center}
\small
\setlength{\tabcolsep}{3.5pt} 
\captionof{table}{Conditional states $A^i_{\alpha\beta\gamma}$ corresponding to the Mentor's Bell-basis measurement outcomes.}
\label{tab:conditional}
\vspace{2mm}

\begin{tabular}{ccccccccccc}
\hline
$\alpha$ & $\beta$ & $\gamma$ &
$A^1_{\alpha\beta\gamma}$ &
$A^2_{\alpha\beta\gamma}$ &
$A^3_{\alpha\beta\gamma}$ &
$A^4_{\alpha\beta\gamma}$ &
$A^5_{\alpha\beta\gamma}$ &
$A^6_{\alpha\beta\gamma}$ &
$A^7_{\alpha\beta\gamma}$ &
$A^8_{\alpha\beta\gamma}$ \\
\hline
$\phi$ & $\phi$ & $\phi$ & $000000$ & $100001$ & $010010$ & $110011$ & $001100$ & $101101$ & $011110$ & $111111$ \\
$\phi$ & $\phi$ & $\psi$ & $000001$ & $100000$ & $010011$ & $110010$ & $001101$ & $101100$ & $011111$ & $111110$ \\
$\phi$ & $\psi$ & $\phi$ & $000010$ & $100011$ & $010000$ & $110001$ & $001110$ & $101111$ & $011100$ & $111101$ \\
$\phi$ & $\psi$ & $\psi$ & $000011$ & $100010$ & $010001$ & $110000$ & $001111$ & $101110$ & $011101$ & $111100$ \\
$\psi$ & $\phi$ & $\phi$ & $000100$ & $100101$ & $010110$ & $110111$ & $001000$ & $101001$ & $011010$ & $111011$ \\
$\psi$ & $\phi$ & $\psi$ & $000101$ & $100100$ & $010111$ & $110110$ & $001001$ & $101000$ & $011011$ & $111010$ \\
$\psi$ & $\psi$ & $\phi$ & $000110$ & $100111$ & $010100$ & $110101$ & $001010$ & $101011$ & $011000$ & $111001$ \\
$\psi$ & $\psi$ & $\psi$ & $000111$ & $100110$ & $010101$ & $110100$ & $001011$ & $101010$ & $011001$ & $111000$ \\
\hline
\end{tabular}
\end{center}

Then, Alice performs a Bell-basis measurement on the qubit pair $(a_1,A_1)$ and communicates her measurement outcome to Bob through classical channel. Simultaneously, Bob performs Bell-basis measurements on the qubit pairs $(b_1,B_1)$ and $(b_2,B_2)$ and communicates his measurement outcomes to Alice through classical channel. 

Using the measurement information received from the Mentor and Alice, Bob performs suitable unitary operation on qubit $B_3$, as summarized in Table \ref{tab:bob}, to deterministically recover the state $|\psi_1\rangle$.

\begin{center}
\small
\renewcommand{\arraystretch}{0.95}
\captionof{table}{Recovery operations performed by Bob on qubit $B_3$ to reconstruct $|\psi_1\rangle$.}
\label{tab:bob}
\vspace{2mm}
\begin{tabular}{cccc}
\hline
Mentor's outcome & Alice's outcome & \multirow{2}{*}{Reduced state of $B_3$} & \multirow{2}{*}{Bob's operation} \\
$(m_3,m_6)$      & $(a_1,A_1)$     &                                         &                                 \\
\hline
\multirow{4}{*}{$|\phi^+\rangle$} & $|\phi^+\rangle$ & $x|0\rangle+y|1\rangle$  & $I$ \\
                                  & $|\phi^-\rangle$ & $x|0\rangle-y|1\rangle$  & $\sigma_z$ \\
                                  & $|\psi^+\rangle$ & $y|0\rangle+x|1\rangle$  & $\sigma_x$ \\
                                  & $|\psi^-\rangle$ & $-y|0\rangle+x|1\rangle$ & $i\sigma_y$ \\
\cline{1-4}
\multirow{4}{*}{$|\phi^-\rangle$} & $|\phi^+\rangle$ & $x|0\rangle-y|1\rangle$  & $\sigma_z$ \\
                                  & $|\phi^-\rangle$ & $x|0\rangle+y|1\rangle$  & $I$ \\
                                  & $|\psi^+\rangle$ & $y|0\rangle-x|1\rangle$  & $i\sigma_y$ \\
                                  & $|\psi^-\rangle$ & $y|0\rangle+x|1\rangle$  & $\sigma_x$ \\
\cline{1-4}
\multirow{4}{*}{$|\psi^+\rangle$} & $|\phi^+\rangle$ & $y|0\rangle+x|1\rangle$  & $\sigma_x$ \\
                                  & $|\phi^-\rangle$ & $-y|0\rangle+x|1\rangle$ & $i\sigma_y$ \\
                                  & $|\psi^+\rangle$ & $x|0\rangle+y|1\rangle$  & $I$ \\
                                  & $|\psi^-\rangle$ & $x|0\rangle-y|1\rangle$  & $\sigma_z$ \\
\cline{1-4}
\multirow{4}{*}{$|\psi^-\rangle$} & $|\phi^+\rangle$ & $-y|0\rangle+x|1\rangle$ & $i\sigma_y$ \\
                                  & $|\phi^-\rangle$ & $y|0\rangle+x|1\rangle$  & $\sigma_x$ \\
                                  & $|\psi^+\rangle$ & $x|0\rangle-y|1\rangle$  & $\sigma_z$ \\
                                  & $|\psi^-\rangle$ & $x|0\rangle+y|1\rangle$  & $I$ \\
\hline
\end{tabular}
\end{center}

On the other hand, using the measurement information received from the Mentor and Bob, Alice performs suitable unitary operation on her qubits $(A_2,A_3),$ as summarized in Table \ref{tab:3}, to deterministically recover the state $|\psi_2\rangle$ at her end.

\begingroup
\small
\setlength{\tabcolsep}{3.5pt}
\setlength{\arrayrulewidth}{0.3pt} 
\renewcommand{\arraystretch}{1.15}

\begingroup
\scriptsize
\begin{longtable}{llcccc}
\caption{Recovery operations performed by Alice on qubits $A_2A_3$ to reconstruct $|\psi_2\rangle$.} \label{tab:3} \\
\hline
\multicolumn{2}{l}{Mentor's outcomes} & \multicolumn{2}{c}{Bob's outcomes} & \multirow{2}{*}{Reduced states of $A_1A_2$} & \multirow{2}{*}{Alice's operation} \\
$(m_1,m_4)$ & $(m_2,m_5)$ & $(b_1,B_1)$ & $(b_2,B_2)$ & & \\
\hline
\endfirsthead

\hline
\multicolumn{2}{l}{Mentor's outcomes} & \multicolumn{2}{c}{Bob's outcomes} & \multirow{2}{*}{Reduced states of $A_2A_3$} & \multirow{2}{*}{Alice's operation} \\
$(m_1,m_4)$ & $(m_2,m_5)$ & $(b_1,B_1)$ & $(b_2,B_2)$ & & \\
\hline
\endhead

\endfoot

\hline
\endlastfoot

\multirow{4}{*}{$|\phi^{\pm_4}\rangle$} & \multirow{4}{*}{$|\phi^{\pm_3}\rangle$} 
  & $|\phi^{\pm_1}\rangle$ & $|\phi^{\pm_2}\rangle$ & $\alpha|00\rangle + (\pm_3\pm_2)\beta|10\rangle \pm_1\gamma|01\rangle + (\pm_1\pm_2\pm_3\pm_4)\delta|11\rangle$ & $CZ^{m_4}(\sigma_z^{B_2 \oplus m_5} \otimes \sigma_z^{B_1})$ \\*
  & & $|\psi^{\pm_1}\rangle$ & $|\psi^{\pm_2}\rangle$ & $\alpha|11\rangle + (\pm_3\pm_2)\beta|01\rangle \pm_1\gamma|10\rangle + (\pm_1\pm_2\pm_3\pm_4)\delta|00\rangle$ & $CZ^{m_4}(\sigma_z^{B_2 \oplus m_5}\sigma_x \otimes \sigma_z^{B_1}\sigma_x)$ \\*
  & & $|\phi^{\pm_1}\rangle$ & $|\psi^{\pm_2}\rangle$ & $\alpha|10\rangle + (\pm_3\pm_2)\beta|00\rangle \pm_1\gamma|11\rangle + (\pm_1\pm_2\pm_3\pm_4)\delta|01\rangle$ & $CZ^{m_4}(\sigma_z^{B_2 \oplus m_5}\sigma_x \otimes \sigma_z^{B_1})$ \\*
  & & $|\psi^{\pm_1}\rangle$ & $|\phi^{\pm_2}\rangle$ & $\alpha|01\rangle + (\pm_3\pm_2)\beta|11\rangle \pm_1\gamma|00\rangle + (\pm_1\pm_2\pm_3\pm_4)\delta|10\rangle$ & $CZ^{m_4}(\sigma_z^{B_2 \oplus m_5} \otimes \sigma_z^{B_1}\sigma_x)$ \\
\hline

\multirow{4}{*}{$|\phi^{\pm_4}\rangle$} & \multirow{4}{*}{$|\psi^{\pm_3}\rangle$} 
  & $|\phi^{\pm_1}\rangle$ & $|\phi^{\pm_2}\rangle$ & $\alpha|10\rangle + (\pm_3\pm_2)\beta|00\rangle \pm_1\gamma|11\rangle + (\pm_1\pm_2\pm_3\pm_4)\delta|01\rangle$ & $CZ^{m_4}(\sigma_z^{B_2 \oplus m_5}\sigma_x \otimes \sigma_z^{B_1})$ \\*
  & & $|\psi^{\pm_1}\rangle$ & $|\psi^{\pm_2}\rangle$ & $\alpha|01\rangle + (\pm_3\pm_2)\beta|11\rangle \pm_1\gamma|00\rangle + (\pm_1\pm_2\pm_3\pm_4)\delta|10\rangle$ & $CZ^{m_4}(\sigma_z^{B_2 \oplus m_5} \otimes \sigma_z^{B_1}\sigma_x)$ \\*
  & & $|\phi^{\pm_1}\rangle$ & $|\psi^{\pm_2}\rangle$ & $\alpha|00\rangle + (\pm_3\pm_2)\beta|10\rangle \pm_1\gamma|01\rangle + (\pm_1\pm_2\pm_3\pm_4)\delta|11\rangle$ & $CZ^{m_4}(\sigma_z^{B_2 \oplus m_5} \otimes \sigma_z^{B_1})$ \\*
  & & $|\psi^{\pm_1}\rangle$ & $|\phi^{\pm_2}\rangle$ & $\alpha|11\rangle + (\pm_3\pm_2)\beta|01\rangle \pm_1\gamma|10\rangle + (\pm_1\pm_2\pm_3\pm_4)\delta|00\rangle$ & $CZ^{m_4}(\sigma_z^{B_2 \oplus m_5}\sigma_x \otimes \sigma_z^{B_1}\sigma_x)$ \\
\hline

\multirow{4}{*}{$|\psi^{\pm_4}\rangle$} & \multirow{4}{*}{$|\phi^{\pm_3}\rangle$} 
  & $|\phi^{\pm_1}\rangle$ & $|\phi^{\pm_2}\rangle$ & $\alpha|01\rangle + (\pm_3\pm_2)\beta|11\rangle \pm_1\gamma|00\rangle + (\pm_1\pm_2\pm_3\pm_4)\delta|10\rangle$ & $CZ^{m_4}(\sigma_z^{B_2 \oplus m_5} \otimes \sigma_z^{B_1}\sigma_x)$ \\*
  & & $|\psi^{\pm_1}\rangle$ & $|\psi^{\pm_2}\rangle$ & $\alpha|10\rangle + (\pm_3\pm_2)\beta|00\rangle \pm_1\gamma|11\rangle + (\pm_1\pm_2\pm_3\pm_4)\delta|01\rangle$ & $CZ^{m_4}(\sigma_z^{B_2 \oplus m_5}\sigma_x \otimes \sigma_z^{B_1})$ \\*
  & & $|\phi^{\pm_1}\rangle$ & $|\psi^{\pm_2}\rangle$ & $\alpha|11\rangle + (\pm_3\pm_2)\beta|01\rangle \pm_1\gamma|10\rangle + (\pm_1\pm_2\pm_3\pm_4)\delta|00\rangle$ & $CZ^{m_4}(\sigma_z^{B_2 \oplus m_5}\sigma_x \otimes \sigma_z^{B_1}\sigma_x)$ \\*
  & & $|\psi^{\pm_1}\rangle$ & $|\phi^{\pm_2}\rangle$ & $\alpha|00\rangle + (\pm_3\pm_2)\beta|10\rangle \pm_1\gamma|01\rangle + (\pm_1\pm_2\pm_3\pm_4)\delta|11\rangle$ & $CZ^{m_4}(\sigma_z^{B_2 \oplus m_5} \otimes \sigma_z^{B_1})$ \\
\hline

\multirow{4}{*}{$|\psi^{\pm_4}\rangle$} & \multirow{4}{*}{$|\psi^{\pm_3}\rangle$} 
  & $|\phi^{\pm_1}\rangle$ & $|\phi^{\pm_2}\rangle$ & $\alpha|11\rangle + (\pm_3\pm_2)\beta|01\rangle \pm_1\gamma|10\rangle + (\pm_1\pm_2\pm_3\pm_4)\delta|00\rangle$ & $CZ^{m_4}(\sigma_z^{B_2 \oplus m_5}\sigma_x \otimes \sigma_z^{B_1}\sigma_x)$ \\*
  & & $|\psi^{\pm_1}\rangle$ & $|\psi^{\pm_2}\rangle$ & $\alpha|00\rangle + (\pm_3\pm_2)\beta|10\rangle \pm_1\gamma|01\rangle + (\pm_1\pm_2\pm_3\pm_4)\delta|11\rangle$ & $CZ^{m_4}(\sigma_z^{B_2 \oplus m_5} \otimes \sigma_z^{B_1})$ \\*
  & & $|\phi^{\pm_1}\rangle$ & $|\psi^{\pm_2}\rangle$ & $\alpha|01\rangle + (\pm_3\pm_2)\beta|11\rangle \pm_1\gamma|00\rangle + (\pm_1\pm_2\pm_3\pm_4)\delta|10\rangle$ & $CZ^{m_4}(\sigma_z^{B_2 \oplus m_5} \otimes \sigma_z^{B_1}\sigma_x)$ \\*
  & & $|\psi^{\pm_1}\rangle$ & $|\phi^{\pm_2}\rangle$ & $\alpha|10\rangle + (\pm_3\pm_2)\beta|00\rangle \pm_1\gamma|11\rangle + (\pm_1\pm_2\pm_3\pm_4)\delta|01\rangle$ & $CZ^{m_4}(\sigma_z^{B_2 \oplus m_5}\sigma_x \otimes \sigma_z^{B_1})$ \\
\hline

\end{longtable}
\endgroup

\normalsize

This marks the end of the protocol.

\section{Preparation of Entangled Channels}\label{sec3}

This section is devoted to presenting the preparation of two entangled resources to be used in this protocol.

Start with the six-qubit input state
$$S_0=|0\rangle_{q[0]}\otimes|0\rangle_{q[1]}\otimes|0\rangle_{q[2]}\otimes|0\rangle_{q[3]}\otimes|0\rangle_{q[4]}\otimes|0\rangle_{q[5]}.$$

Apply Hadamard gates $H$ on $q[0],q[1],q[2]$ to transform $S_0$ to 
$$S_1=\frac{1}{\sqrt{2}}(|0\rangle+|1\rangle)_{q[0]}\otimes\frac{1}{\sqrt{2}}(|0\rangle+|1\rangle)_{q[1]}\otimes\frac{1}{\sqrt{2}}(|0\rangle+|1\rangle)_{q[2]}\otimes|0\rangle_{q[3]}\otimes|0\rangle_{q[4]}\otimes|0\rangle_{q[5]}.$$

Apply CNOT gates to the qubit pairs $(q[0],q[5]),(q[1],q[4]),(q[2],q[3])$ to transform $S_1$ to the first entangled channel:
\begin{equation*}
\begin{split}
|\Omega_1\rangle_{M-A} &= \frac{1}{2\sqrt{2}} \Big( |000000\rangle + |001100\rangle + |010010\rangle + |011110\rangle \\
&\qquad\qquad + |100001\rangle + |101101\rangle + |110011\rangle + |111111\rangle \Big)_{q[0]q[1]q[2]q[3]q[4]q[5]}
\end{split}
\end{equation*}

Moreover, if a SWAP gate is applied to the qubit pair $(q[0],q[2])$, then $|\Omega_1\rangle_{M-A}$ transforms to the second entangled channel:
\begin{equation*}
\begin{split}
|\Omega_2\rangle_{M-B} &= \frac{1}{2\sqrt{2}} \Big( |000000\rangle + |001001\rangle + |010010\rangle + |011011\rangle \\
&\qquad\qquad + |100100\rangle + |101101\rangle + |110110\rangle + |111111\rangle \Big)_{q[0]q[1]q[2]q[3]q[4]q[5]}
\end{split}
\end{equation*}

  \begin{figure}[H]
    \centering

    \begin{subfigure}[b]{0.345\textwidth}
        \centering
        \includegraphics[width=\textwidth]{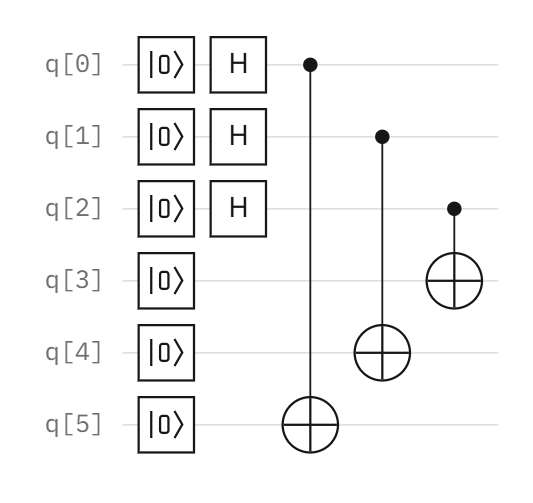}
        \caption{Quantum circuit for $|\Omega_1\rangle_{M-A}$}
        \label{fig:phi1}
    \end{subfigure}
    \hfill
    \begin{subfigure}[b]{0.40\textwidth}
        \centering
        \includegraphics[width=\textwidth]{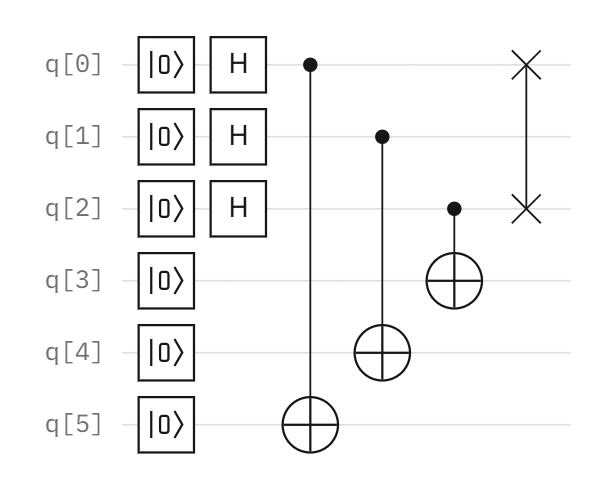}
        \caption{Quantum circuit for $|\Omega_2\rangle_{M-B}$}
        \label{fig:phi2}
    \end{subfigure}

    \caption{Quantum circuit for the preparation of entangled channels.}
    \label{fig:entangled_channels}
\end{figure}

\section{Simulation in Qiskit Environment}\label{sec4}

This section is devoted to examining the feasibility of the protocol designed in the Section \ref{sec2}. In what follows, we design a quantum circuit corresponding to the protocol and execute it using Qiskit’s “AerSimulator” and analyze the outcomes. In this circuit (Fig. \ref{fig2}), qubits $q[0],$ $q[1],$ $q[2],$ $q[3],$ $q[4],$ $q[5],$ $q[6],$ $q[7],$ $q[8],$ $q[9],$ $q[10],$ $q[11],$ $q[12],$ $q[13]$ and $q[14]$ represent the qubits $a_1,b_1,b_2,m_1,m_2,m_3,A_1,A_2,A_3,m_4,m_5,m_6,B_1,B_2$ and $B_3$ respectively. Since Bell-basis measurements cannot be directly realized in Qiskit, we use a combination of Hadamard and CNOT gates followed by computational-basis measurements to realize measurements in the Bell basis. A stepwise description of circuit construction (Fig. \ref{fig2}) is given below:

\begin{itemize}

\item\textsc{Part I:} This part involves the preparation of the initial qubits $|\psi_{1}\rangle$ and $|\psi_{2}\rangle$.

\item\textsc{Part II:} In this part, the entangled channel $|\Omega_{1}\rangle_{M-A}$ is prepared.

\item\textsc{Part III:} In this part, the entangled channel $|\Omega_{2}\rangle_{M-B}$ is prepared.

\item\textsc{Part IV:} In this part, the mentor performs a measurement on the qubit pair $(m_{1}, m_{4})$ in the Bell basis.

\item\textsc{Part V:} This part involves the mentor performing a measurement on the qubit pair $(m_{2}, m_{5})$ in the Bell basis.

\item\textsc{Part VI:} In this part, the mentor performs a measurement on the qubit pair $(m_{3}, m_{6})$ in the Bell basis.

\item\textsc{Part VII:} In this part, Alice performs a Bell-basis measurement on her qubits $a_{1}$ and $A_{1}.$

\item\textsc{Part VIII:} This part involves Bob performing a Bell-basis measurement on his qubits $b_{1}$ and $B_{1}.$

\item\textsc{Part IX:} In this part, Bob performs a Bell-basis measurement on his qubits $b_{2}$ and $B_{2}$

\item\textsc{Part X:} This part of the circuit implements the unitary operations performed by Alice on her qubits $A_{2}$ and $A_{3}$ to reconstruct the state $|\psi_{2}\rangle$, which depend on the measurement results of the mentor and Bob

\item\textsc{Part XI:} This part of the circuit implements the unitary operations performed by Bob on his qubit $B_{3}$ to recover the state $|\psi_{1}\rangle$, based on the measurement results of the mentor and Alice.
\end{itemize}

\begin{figure}[H]
    \centering
    \includegraphics[width=1\linewidth]{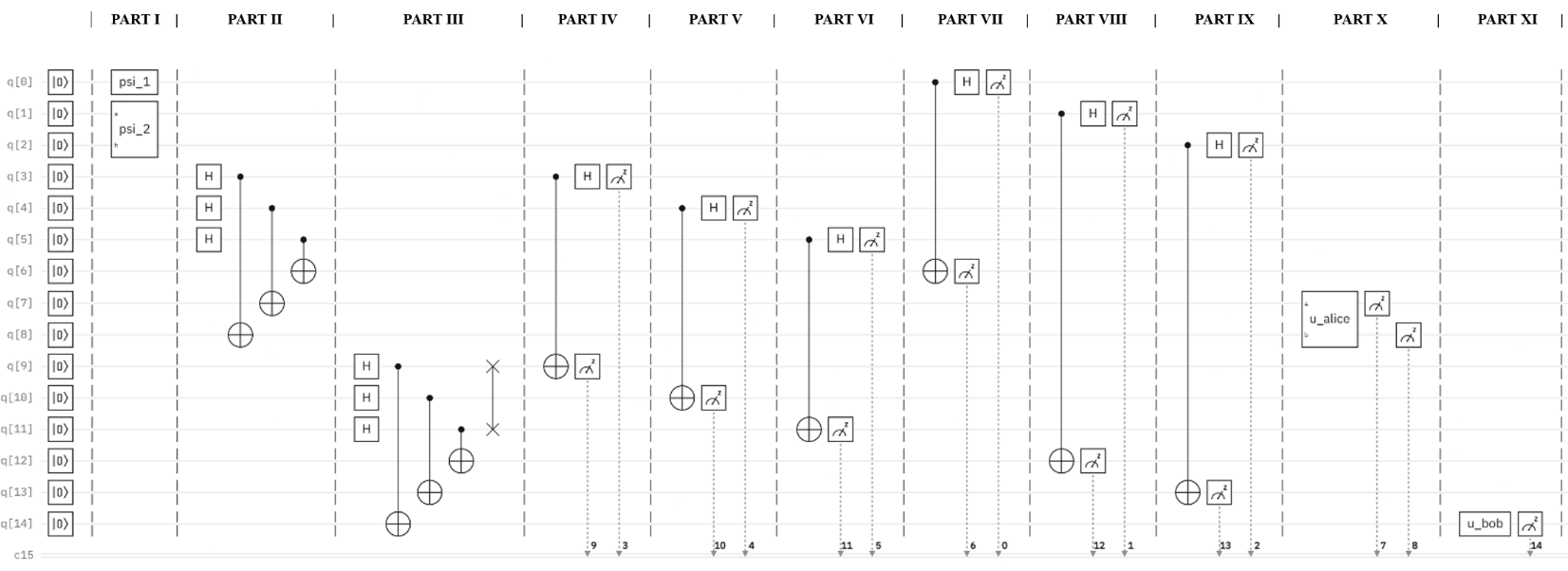}
    \caption{Circuit diagram for the proposed protocol}\label{fig2}
\end{figure}

To demonstrate the protocol, we initialize the circuit using specific arbitrary states for $|\psi_1\rangle_{a_1}$ and $|\psi_2\rangle_{b_1b_2}$ to analyze the final outcomes. 

We set the single-qubit state to
$$|\psi_1\rangle_{a_1}=\frac{\sqrt{3}}{2}|0\rangle+e^{\frac{i\pi}{2}}\frac{1}{2}|1\rangle,$$
which can be prepared from the initial $|0\rangle$ state using the following gate sequence:
$$|\psi_1\rangle_{a_1}=P\left(\frac{\pi}{2}\right)R_Y\left(\frac{\pi}{3}\right)P\left(\frac{\pi}{2}\right)|0\rangle.$$

For the two-qubit state, we choose
$$|\psi_2\rangle_{b_1b_2}=\frac{1}{2}\left(e^{i\frac{7\pi}{4}}|00\rangle+e^{i\frac{\pi}{4}}|01\rangle+e^{i\frac{7\pi}{4}}|10\rangle+e^{i\frac{5\pi}{4}}|11\rangle\right),$$
which is generated from the $|00\rangle$ state through the transformation:
$$|\psi_2\rangle_{b_1b_2}=\left(I\otimes R_Z\left(\frac{\pi}{2}\right)\right) \left(I\otimes H\right) \left(CNOT_{b_2\to b_1}\right) \left(I\otimes R_Y\left(\frac{\pi}{2}\right)\right) |00\rangle.$$

Although Section \ref{sec2} requires no post-correction measurements, our Qiskit simulation (using the \texttt{AerSimulator} backend) includes computational-basis measurements (Parts X and XI) strictly to validate the recovered states. After execution, the target qubits at Alice's and Bob's terminals are expected to match the initial input states: $$ |\psi_2\rangle_{A_2A_3} = \frac{1}{2}\left(e^{i\frac{7\pi}{4}}|00\rangle+e^{i\frac{\pi}{4}}|01\rangle+e^{i\frac{7\pi}{4}}|10\rangle+e^{i\frac{5\pi}{4}}|11\rangle\right) \quad \text{and} \quad |\psi_1\rangle_{B_3} = \frac{\sqrt{3}}{2}|0\rangle+e^{\frac{i\pi}{2}}\frac{1}{2}|1\rangle. $$

To evaluate performance, we executed the circuit for $30\,000$, $50\,000$, and $100\,000$ shots. The resulting histograms show consistent probability distributions across all sample sizes, confirming the protocol's reliability. These specific input states are purely illustrative; the proposed mechanism remains universally applicable to any arbitrary quantum states.

\begin{figure}[H]
    \centering
    \includegraphics[width=0.75\linewidth]{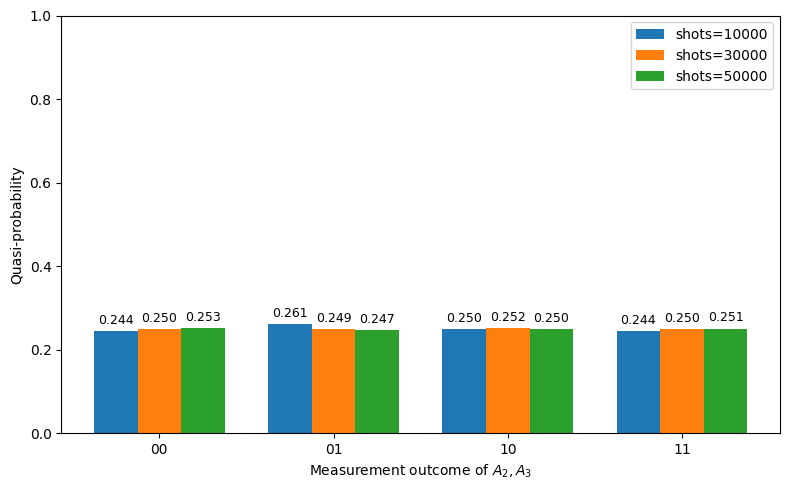}
    \caption{Probability distribution of the final output states $A_2A_3$}\label{fig3}
\end{figure}

\begin{figure}[H]
    \centering

\includegraphics[width=0.75\linewidth]{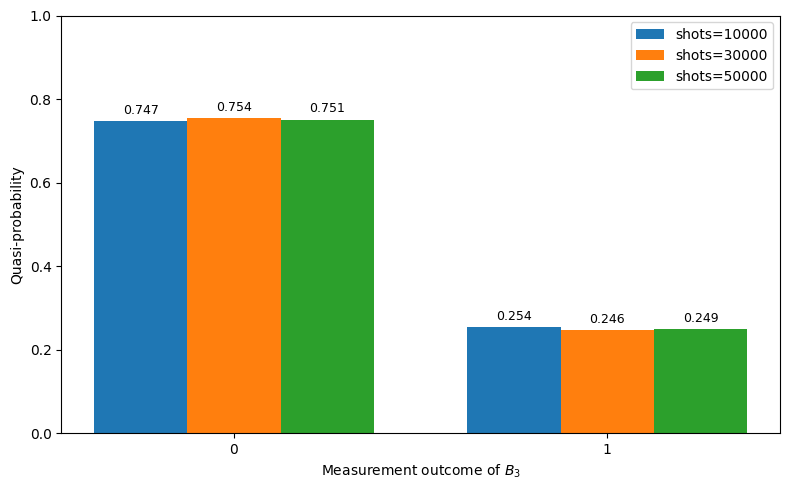}
\caption{Probability distribution of the final output states $B_3$}\label{fig4}
\end{figure}

The simulation results shown in Fig. \ref{fig3} and Fig. \ref{fig4} yield probability distributions that are very close to the theoretical probabilities of the corresponding target states. In particular, the four computational-basis outcomes of $A_2A_3$ have simulated probabilities very close to the theoretical value of $25\%$ for each outcome, while the simulated probabilities of obtaining $0$ and $1$ at $B_3$ are very close to the theoretical values of $75\%$ and $25\%$, respectively. These results indicate that the simulation captures the expected probability distributions of the output states and hence provides numerical evidence for the successful reconstruction of the intended states at Alice's and Bob's terminals.

\section{Efficiency of the Protocol}\label{sec5}

To assess the performance and resource footprint of the proposed scheme, we evaluate its intrinsic communication efficiency. A standard metric used to quantify the efficiency of a quantum teleportation protocol is given by the relation
\begin{equation}
    \eta = \frac{q_t}{q_c + b_t},
\end{equation}
where $q_t$ denotes the total number of qubits successfully teleported, $q_c$ represents the total number of entangled qubits consumed from the shared quantum resources, and $b_t$ indicates the total classical communication cost measured in classical bits.

In the context of the present protocol, the relevant parameters are determined as follows:
\begin{itemize}
    \item The total number of teleported qubits is $q_t = 3$, as Alice transmits an arbitrary single-qubit state and Bob simultaneously transmits an arbitrary two-qubit state.
    \item The protocol utilizes two separate six-qubit entangled channels ($|\Omega_1\rangle_{M-A}$ and $|\Omega_2\rangle_{M-B}$), resulting in a total quantum resource consumption of $q_c = 12$ qubits.
    \item The classical communication cost is accrued through the transmission of measurement outcomes across the network. The mentor performs three Bell-basis measurements on two-qubit pairs, generating $3 \times 2 = 6$ classical bits. Alice performs one Bell-basis measurement, generating $2$ classical bits. Bob performs two Bell-basis measurements, generating $2 \times 2 = 4$ classical bits. Consequently, the aggregate classical communication cost is $b_t = 6 + 2 + 4 = 12$ classical bits.
\end{itemize}

Substituting these values into the efficiency equation yields:
\begin{equation}
    \eta = \frac{3}{12 + 12} = \frac{3}{24} = 12.5\%.
\end{equation}

For comparison, while the theoretical efficiency of our proposed protocol ($12.5\%$) is slightly lower than the mentor-initiated asymmetric bidirectional teleportation protocol proposed in \cite{mentorBQT}, that achieved an efficiency of approximately $13.64\%$, this specific trade-off is justified by the broader capability of the scheme. Our protocol is designed to be universally applicable for the teleportation of all arbitrary two-qubit states, whereas the architecture in \cite{mentorBQT} is restricted to specific two-qubit Bell-like states.

\section{Discussion and Conclusion}\label{sec6}

In this paper, we proposed a mentor-initiated asymmetric bidirectional quantum teleportation protocol for the simultaneous exchange of an arbitrary single-qubit state from Alice to Bob and an arbitrary two-qubit state from Bob to Alice. A key advantage of our scheme is its universal applicability to arbitrary two-qubit states, whereas recent mentor-based protocol is restricted to specific Bell-like states \cite{mentorBQT}. Furthermore, our protocol is operationally straightforward, relying exclusively on standard bipartite Bell-basis measurements and completely avoiding the need for complex multi-qubit joint measurements. The theoretical framework was successfully validated through quantum circuit simulations using Qiskit, confirming the feasibility and robustness of the state exchange.

Looking ahead, while the current protocol operates efficiently between the primary nodes, introducing a supervisory authorization mechanism could enhance its utility in secure quantum networks. Therefore, integrating one or more controller to supervise the bidirectional state exchange process represents a highly promising direction for future research.


\begin{thebibliography}{99}

\small

\bibitem{Aisan2025}
Y. Aisan, N. Maihemuti, J. Peng, and J. Tang,
``Conclusive bidirectional quantum teleportation of arbitrary unknown single-particle states in amplitude-damping channel,''
\textit{International Journal of Quantum Information}, vol. 23, no. 05, 2550008, 2025. 

\bibitem{Bennett1993}
C. H. Bennett, G. Brassard, C. Cr\'epeau, R. Jozsa, A. Peres, and W. K. Wootters,
``Teleporting an unknown quantum state via dual classical and Einstein-Podolsky-Rosen channels,''
\textit{Physical Review Letters}, vol. 70, no. 13, p. 1895, 1993. 

\bibitem{Choudhury2019}
B. S. Choudhury and S. Samanta,
``A remote state preparation scheme initiated and fixed by a Mentor,''
\textit{Physics of Particles and Nuclei Letters}, vol. 16, no. 6, pp. 608--612, 2019. 

\bibitem{Choudhury2022}
B. S. Choudhury and S. Samanta,
``A mentor-initiated joint remote state preparation scheme for qubits,''
\textit{Pramana}, vol. 96, no. 1, p. 9, 2022. 

\bibitem{Choudhury2023a}
B. S. Choudhury, M. K. Mandal, S. Samanta, and B. Dolai,
``A mentor initiated bi-directional controlled remote state preparation protocol for non-maximally entangled Bell and GHZ states,''
\textit{Chinese Journal of Physics}, vol. 84, pp. 27--38, 2023. 

\bibitem{Choudhury2023b}
B. S. Choudhury, M. K. Mandal, B. Dolai, and S. Samanta,
``Bi-directional controlled asymmetric teleportation protocol initiated by a mentor in noisy environments,''
\textit{Physica Scripta}, vol. 98, no. 9, 095107, 2023. 

\bibitem{Choudhury2026}
B. S. Choudhury and S. Samanta,
\textit{Mathematical Theory of Quantum Teleportation Protocols: An Introduction for Scientists and Engineers},
CRC Press, 2026.

\bibitem{Hong2016}
W. Q. Hong,
``Asymmetric bidirectional controlled teleportation by using a seven-qubit entangled state,''
\textit{International Journal of Theoretical Physics}, vol. 55, no. 1, pp. 384--387, 2016. 

\bibitem{Jing2024a}
R. H. Jing, Y. B. Huang, A. A. Bi, W. W. Luo, P. Zhou, and Q. Lan,
``Mentor initiated multiparty hierarchical joint remote preparation of an arbitrary $n$-qudit state via generalized Bell states,''
\textit{Physica Scripta}, vol. 99, no. 2, 025103, 2024. 

\bibitem{Li2019}
Y. H. Li, Y. Qiao, M. H. Sang, and Y. Y. Nie,
``Bidirectional controlled remote state preparation of an arbitrary two-qubit state,''
\textit{International Journal of Theoretical Physics}, vol. 58, no. 7, pp. 2228--2234, 2019. 

\bibitem{mentorBQT}
P. Saha, B. S. Choudhury, B. Tripathy, and S. Das,
``Mentor initiated bi-directional asymmetric quantum teleportation protocol with hierarchical control,''
\textit{Modern Physics Letters A}, vol. 41, no. 09, 2650035, 2026. 

\bibitem{Vaidman1994}
L. Vaidman,
``Teleportation of quantum states,''
\textit{Physical Review A}, vol. 49, no. 2, p. 1473, 1994. 

\bibitem{Wang2022}
M. Wang and H. S. Li,
``Bidirectional quantum teleportation using a five-qubit cluster state as a quantum channel,''
\textit{Quantum Information Processing}, vol. 21, no. 2, p. 44, 2022. 

\bibitem{Yang2016}
Y. Q. Yang, X. W. Zha, and Y. Yu,
``Asymmetric bidirectional controlled teleportation via seven-qubit cluster state,''
\textit{International Journal of Theoretical Physics}, vol. 55, no. 10, pp. 4197--4204, 2016. 

\bibitem{Zha2013}
X. W. Zha, Z. C. Zou, J. X. Qi, and H. Y. Song,
``Bidirectional quantum controlled teleportation via five-qubit cluster state,''
\textit{International Journal of Theoretical Physics}, vol. 52, no. 6, pp. 1740--1744, 2013. 

\bibitem{Zhang2015}
D. Zhang, X. W. Zha, and Y. J. Duan,
``Bidirectional and asymmetric quantum controlled teleportation,''
\textit{International Journal of Theoretical Physics}, vol. 54, no. 5, pp. 1711--1719, 2015. 



\end{thebibliography}
\end{document}